\documentclass[sigconf]{acmart}

\AtBeginDocument{%
  \providecommand\BibTeX{{%
    \normalfont B\kern-0.5em{\scshape i\kern-0.25em b}\kern-0.8em\TeX}}}

\copyrightyear{2021}
\acmYear{2021}
\setcopyright{iw3c2w3}
\acmConference[WWW '21]{Proceedings of the Web Conference 2021}{April 19--23, 2021}{Ljubljana, Slovenia}
\acmBooktitle{Proceedings of the Web Conference 2021 (WWW '21), April 19--23, 2021, Ljubljana, Slovenia}
\acmPrice{}
\acmDOI{10.1145/3442381.3449791}
\acmISBN{978-1-4503-8312-7/21/04}

\setcopyright{iw3c2w3} 
\begin{document}

\title{Future-Aware Diverse Trends Framework for Recommendation}


\author{Yujie Lu}
\affiliation{%
  \institution{Tencent}
  \city{Shenzhen}
  \state{Guangdong}
  \country{China}
}
\email{yujielu10@gmail.com}

\author{Shengyu Zhang}
\authornote{Both authors contributed equally to this research.}
\affiliation{%
  \institution{Zhejiang University}
  \city{Hangzhou}
  \state{Zhejiang}
  \country{China}}
\email{sy\_zhang@zju.edu.cn}
\author{Yingxuan Huang}
\authornotemark[1]
\affiliation{%
  \institution{Tencent}
  \city{Shenzhen}
  \state{Guangdong}
  \country{China}}
\email{eloisehuang.yx@gmail.com}

\author{Luyao Wang}
\affiliation{%
  \institution{Zhejiang University}
  \city{Hangzhou}
  \state{Zhejiang}
  \country{China}}
\email{luyaowang@zju.edu.cn}

\author{Xinyao Yu}
\affiliation{%
  \institution{Zhejiang University}
  \city{Hangzhou}
  \state{Zhejiang}
  \country{China}}
\email{yuxinyao@zju.edu.cn}

\author{Zhou Zhao}
\authornote{Corresponding Authors.}
\affiliation{%
  \institution{Zhejiang University}
  \city{Hangzhou}
  \state{Zhejiang}
  \country{China}}
\email{zhaozhou@zju.edu.cn}
\author{Fei Wu}
\authornotemark[2]
\affiliation{%
  \institution{Zhejiang University}
  \city{Hangzhou}
  \state{Zhejiang}
  \country{China}}
\email{wufei@zju.edu.cn}


\renewcommand{\shortauthors}{Yujie Lu, et al.}
\newcommand{\etal}{\textit{et al}.}
\newcommand{\ie}{\textit{i}.\textit{e}.}
\newcommand{\eg}{\textit{e}.\textit{g}.}

\begin{abstract}
In recommender systems, modeling user-item behaviors is essential for user representation learning. Existing sequential recommenders consider the sequential correlations between historically interacted items for capturing users' historical preferences. However, since users' preferences are by nature time-evolving and diversified, solely modeling the historical preference (without being aware of the time-evolving trends of preferences) can be inferior for recommending complementary or fresh items and thus hurt the effectiveness of recommender systems. In this paper, we bridge the gap between the past preference and potential future preference by proposing the future-aware diverse trends (FAT) framework. By \texttt{future-aware}, for each inspected user, we construct the future sequences from other similar users, which comprise of behaviors that happen after the last behavior of the inspected user, based on a proposed neighbor behavior extractor. By \texttt{diverse trends}, supposing the future preferences can be diversified, we propose the diverse trends extractor and the time-aware mechanism to represent the possible trends of preferences for a given user with multiple vectors. We leverage both the representations of historical preference and possible future trends to obtain the final recommendation. The quantitative and qualitative results from relatively extensive experiments on real-world datasets demonstrate the proposed framework not only outperforms the state-of-the-art sequential recommendation methods across various metrics, but also makes complementary and fresh recommendations.

\end{abstract}




\maketitle

\begin{figure}[htb]
    \centering
    \includegraphics[width=\linewidth]{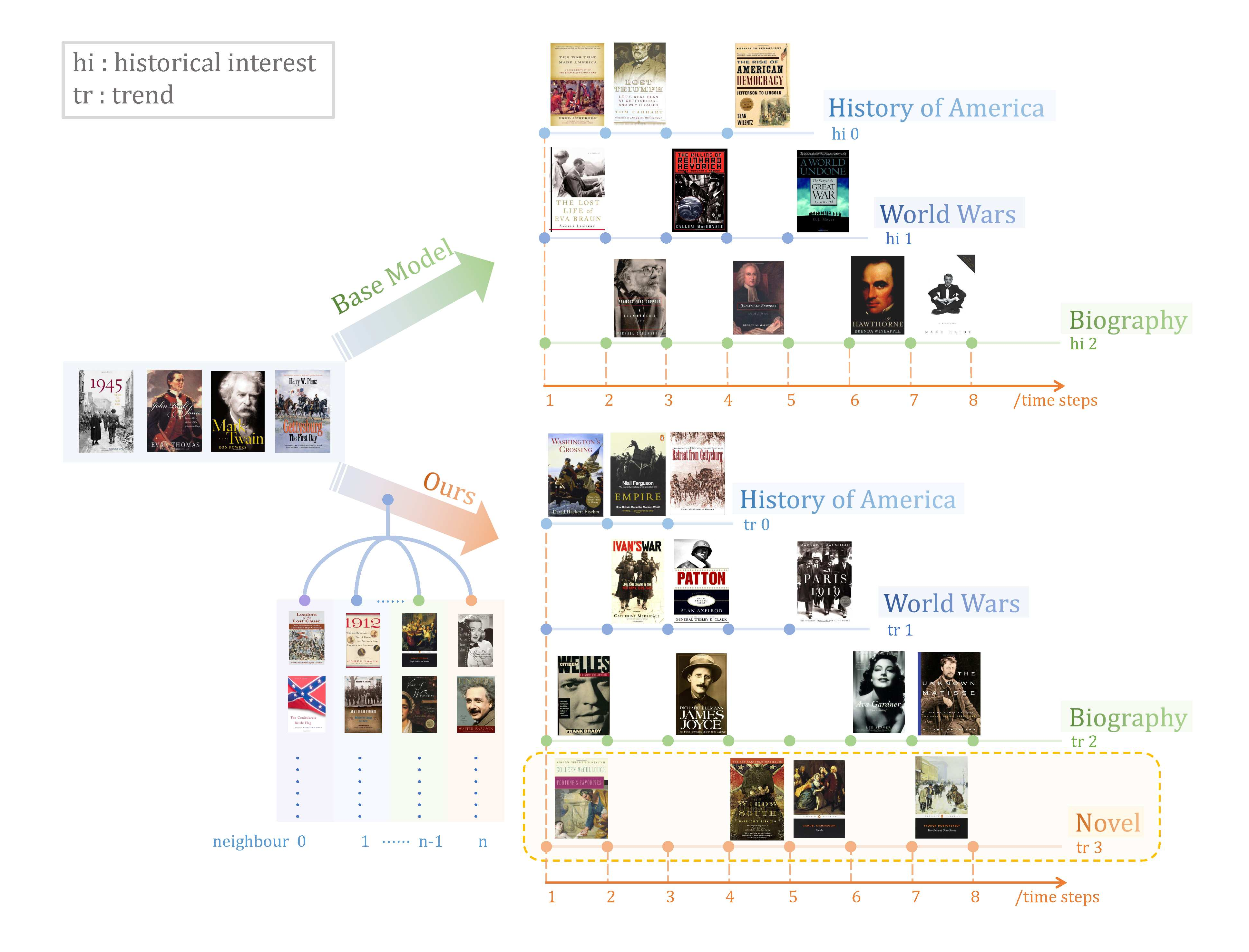}
    \caption{A diagram of our model and base model. On the left the light blue sequence is the historical sequence clicked by the user. On the right the labels such as "History of America" indicate categories of items, while the part highlighted in the yellow box below is fresh items whose types can't be found in the historical sequence but can be inferred from trends.} 
    \label{fig:trendCompare}
\end{figure}

\section{Introduction}
Recommender systems assume a central part of many real-world applications (\eg, e-commerce platforms \cite{Zhang_Tan_Zhao_Yu_Kuang_Jiang_Zhou_Yang_Wu_2020,Wang_Huang_Zhao_Zhang_Zhao_Lee_2018,Zhang_Tan_Yu_Zhao_Kuang_Liu_Zhou_Yang_Wu_2020}) with the prevalence of the Internet and information technology. On online platforms, users interact with a series of items in a chronological order, implying continuous and temporary correlation between each item. In this scenario, the sequential recommenders have become indispensable techniques in the recommendation area, which aim at predicting the next item that the user may interact with by modeling users' preferences on the basis of sequential dependencies among the users' historical interactions.

Existing sequential recommendation algorithms model and represent user preferences in  various manners. 
Most conventional models such as Markov chain-based \cite{rendle2010FPMC,he2016Vista,he2017translation} and factorization-based \cite{rendle2011contextaware,kabbur2013fism,zhe2015usertopic} ones have successfully captured users' short-term and long-term interests by adopting Markov chains and matrix factorization respectively, but either failed to model intricate dynamics or ignored the time dependency. In contrast, deep learning based techniques typically represent user preferences with low-dimensional embedding vectors. For instance, the deep neural network proposed for YouTube video recommendation(YouTubeDNN) \cite{paul2016youtubednn} represents each user by one fixed-length vector transformed from the past practices of users, while not appropriate for modeling various interests because of the dimension explosion. To tackle this issue, Deep Interest Network (DIN)\cite{zhou2017din} makes the user representation vary over different items with attention mechanisms to capture the diversity of user interests. More recent works \cite{li2019mind,cen2020cmind} propose to encode users' historical behaviors into users' varying interests by leveraging capsule routing mechanism. Nevertheless, all of these methods model user preferences only taking into account the past behaviors of users, ignoring the potential future preference and failing to capture the time-evolving trends of user diversified preferences. 

We argue that preferences changing over time of similar users is an extra important factor to model future diverse preference trends in addition to historical preference. Such trends can be summarized from the relative future behaviors of users with similar interests. Specifically, for an inspected user, other users with similar interests tends to have common behaviors (e.g. click the same items) with the inspected user, and the behaviors that happen after the common behavior can be viewed as the relative future behaviors. As shown in \ref{fig:trendCompare}, models that solely focus on users' historical interests tend to recommend similar and complementary items. In contrast, the proposed future-aware diverse trends framework is capable of recommending fresh items that may seem irrelevant to the user's historical preference, while in reality are consistent with one of the preference trends that can be captured from the future behaviors of users' with similar interest (or at least similar behaviors).

In this paper, we focus on the problem of modeling diverse trends of users for sequential recommendation. In order to overcome the limitations of existing methods, we propose the future-aware diverse trends (FAT) framework for learning user representations that reflect diverse trends of users preferences. To infer the user representation vectors, we design an implicit neighbor behaviors extractor(INBE) layer and a novel diverse trends capture layer. To construct neighborhoods implicitly, the INBE module utilizes Pearson Correlation Coefficient \cite{breese2013empirical} and an interaction-based users filter. The diverse trends capture layer applies dynamic routing and time-aware mechanism to adaptively aggregate neighbor user's relative future behaviors as user trend representation. The user representation is then computed by concatenating the user historical behaviors embedding from traditional sequence modeling and the user trends embedding. The process of dynamic routing can be viewed as soft-clustering, which groups similar users' relative future behaviors into several clusters. Each cluster of future behaviors is further used to infer the user trend representation vector according to the time-varying attention of each trend corresponding to the specific items. In this way, for a particular user, FAT outputs the final user preference representations considering both the user past preference and potential future preference. To summarize, the main contributions of this work are as follows:

\begin{itemize}
    \item To better infer the dynamics of user behaviors, we design a FAT framework, which leverage the future information and capture diverse trends of user preference.
    \item We first design a neighbor behavior module to extract relative future behaviors from similar users implicitly. We design the diverse trends capture module, which utilizes dynamic routing to adaptively aggregate neighbor's future behaviors into trend representation vectors. We then leverage time-aware mechanism over trends to better model time-varying user potential preferences.
    \item Compared with existing methods, FAT shows superior performance on several public datasets over metrics such as Recall and NDCG. In addition, we conduct experiments to show that FAT can bring diversity of retrieved items better than other baselines.
\end{itemize}

The remainder of this paper is organized as follows: related works are reviewed in Section \ref{sec:related}; Section \ref{sec:methodology} formulates the sequential recommendation task and elaborates the technical details of FAT; In Section \ref{sec:EXP}, we detail the experiments for comparing FAT with existing methods on several public benchmarks; The last section gives conclusion and future work of this paper.

\section{Related Work}
\label{sec:related}
\subsection{Sequential Recommendation} 

Conventional sequential recommendation popular models usually use matrix factorization and Markov chains to capture long-term and short-term interests of users, respectively. The Markov chain-based sequential recommendation algorithms use functions obtained from past transactions to predict the user's next interaction. Personalized Markov Chain Factorization (FPMC) \cite{rendle2010FPMC} combines the advantages of Markov Chain and Matrix Factorization. Since the operation used is linear, FPMC cannot capture the interaction between multiple factors, because each component independently affects the user's next interaction. Hierarchical Representation Model (HRM) solves this problem by summarizing multiple interaction factors through nonlinear maximum pooling operations\cite{wang2015HRM}. HRM uses continuous value representations of users and items, and builds a mixed representation on users and items based on previous interactions. Both FMPC and HRM only model local interactions between successive transactions.

Due to the strong representation learning capability \cite{Zhang_Jiang_Wang_Kuang_Zhao_Zhu_Yu_Yang_Wu_2020}, deep learning techniques have also been adopted in the sequential recommendation in recent years. DREAM \cite{yu2016Dream}, based on Recurrent Neural Network (RNN), learns the user's dynamic representation to reveal the user's dynamic interest. DIN \cite{zhou2017din} designs a local activation unit to adaptively learn the representation of user interests from past behaviors with respect to a certain ad. \cite{yuan2020future} proposes a encoder-decoder networks to integrate future data into model training. \cite{guo2020IMFOU} propose to model user intention from both ordered and unordered facets simultaneously. Contextualized Temporal Attention Mechanism proposed in \cite{wu2020deja} learns to weigh historical actions' influence considering different contexts.

\subsection{User Modelling}
Representing users as vectors is commonly used in recommender system. Traditional methods assembles user preference as vectors composed of interested items \cite{bell2007neighborCF,herlocker2002EmpNeighborCF,sarwar2001itemCF}, keywords \cite{cantador2010contentRec,elkahky2015multiViewUM} and topics \cite{Yin2015dUM}. As the emergence of distributed representation learning, user embeddings obtained by neural networks are widely used. \cite{chen2016UPdistributedRep} employs RNN-GRU to learn user embeddings from the temporal ordered review documents. \cite{rakkappan2019stackedrnn} utilizes Stacked Recurrent Neural Networks to capture the evolution of contexts and temporal gaps. \cite{fan2019graph} proposes the framework GraphRec to jointly capture interactions and opinions in the user-item graph.

GRU4Rec \cite{hidasi2015gru4rec} introduces recurrent neural networks for the recommender systems firstly. \cite{Bogina2017dwell,hidasi2018topKgains,kiam2016improvedRNN,YaoDongZhangAAAI2021} models behavior sequence. \cite{kiam2016improvedRNN} applies data augmentation to enhance training of GRU4Rec. \cite{Bogina2017dwell} considers the dwell time. \cite{hidasi2018topKgains} provides impressive top-k gains for recurrent neural networks for session-based recommendation with a proposed new class of loss functions coupled with an additional sampling (combination of uniform sampling and popularity sampling) for negative sampling in GRU4Rec.
\cite{hidasi2016parallelRNN} considers additional item information other than IDs(parallel RNN). \cite{jan2017neighbor} combines the session-based KNNs with GRU4Rec using the methods of switching, cascading, and weighted hybrid.

\cite{yuyu2014ctr} proposes a RNN-based framework for click-through rate(CTR) prediction in sponsor search. RRN \cite{wu2017RRN} is the first recurrent recommender network that attempts to capture the dynamics of both user and item representation. \cite{bharadhwaj2018explanations} further improves the RRN's interpretability by devising a time-varying neighborhood style explanation scheme. \cite{chen2018memory} proposes a memory-augmented neural network for the sequential recommendation, with analogous gains observed in other domains \cite{Duan_Tang_Zhang_Zhang_Zhao_Xue_Zhuang_Wu_2018,Zhang_Dong_Hu_Guo_Wu_Xie_Wu_2018,Duan_Zhang_Zhao_Wu_Zhuang_2018}. \cite{Quadrana2017HRNN, donkers2017userbasedRNN} use GRU to model users and sessions. \cite{devooght2017LSTR} uses RNN for the collaborative filtering task and considered two different objective functions in the RNN model. \cite{mass2017intersession} deploys a multi-layer GRU network to capture sequential dependencies and user interest from both the inter-session and intra-session levels. HNVM \cite{xiao2019hier} models different levels of user preferences via a unified hierarchical generative process. 

NextItNet \cite{yuan2018simple} is a generative CNN model with the residual block structure for the sequential recommendation. RCNN proposed in \cite{xu2019RCNN} utilizes the recurrent architecture of RNN and the convolutional operation of CNN to extract long-term and short-term patterns respectively.

\section{Methodology}
\label{sec:methodology}
In this section, we first formulate the sequential recommendation problem, then introduce the proposed framework in detail. We lastly discuss the prediction and network training procedure of FAT.
\subsection{Problem Formulation}
In a typical recommendation scenario, we have a set of users and a set of items which can be denoted as ${U=\{u_1, u_2, ..., u_{|U|}\}}$ and ${V=\{v_1, v_2, ..., v_{|V|}\}}$, respectively. Let ${X_u = \{ x_1^u, x_2^u, ..., x_{|X_u|}^u \}}$ denote the sequence of interacted items from user ${u \in U}$ sorted in a chronological order: ${x_t^u}$ denotes the item that the user ${u}$ has interacted with at time step ${t}$. Given the user historical behaviors, the goal of the sequential recommendation task considered in this paper is to retrieve a subset of items from the pool ${V}$ for each user in ${U}$ such that the user is most likely to interact with the recommended items. Notations are summarized in \ref{tab:notation}.

\begin{table}[h]
\caption{Notations.}
\centering
\resizebox{\columnwidth}{!}{%
\begin{tabular}{l l}
\toprule
Notation & Description \\

    \midrule
    u   & a user \\
    v   & an item \\ 
    x   & an interaction \\ 
    d   & the dimension of user/item embeddings \\
    t   & the number of trends \\
    U   & the set of users \\
    V   & the set of items \\
    X   & the set of interactions \\
    T   & the trends set \\
    N   & the number of retrieved items \\
    \bottomrule
\end{tabular}%
}
    \label{tab:notation}
\end{table}

Specifically, each instance is represented by a tuple ${(X_u, T_u, A_i)}$, where ${X_u}$ denotes the set of items interacted by user ${u}$, ${T_u}$ denotes the relative future sequence set extracted from similar users, detail will be illustrated in the Section \ref{sec:INBE}, ${A_i}$ the features of target item ${i}$ including the information of interaction time and item ID.

To model diverse user preferences dynamically, FAT learns a function ${f}$ for mapping user's corresponding interactions ${X_u}$ and trend set ${T_u}$ into user representations, which can be formulated as
\begin{equation}
    \overrightarrow{e_u} = f(X_u, T_u)
\label{eq:userRepFunc}
\end{equation}
where ${\overrightarrow{e_u} \in \mathbb{R}^{d\times1}}$ denotes the representation vector of user u, d the dimension. Besides, the representation vector of target item ${i}$ is obtained by an embedding function ${g}$ as
\begin{equation}
    \overrightarrow{e_i} = g(A_i)
\label{eq:itemRepFunc}    
\end{equation}
where ${\overrightarrow{e_i} \in \mathbb{R}^{d\times1}}$ denotes the representation vector of item i, and the detail of ${g}$ will be illustrated in the Section \ref{sec:seqMod}.

When user representation vector and item representation vector are learned, top-N items are recommended according to the likelihood function ${p}$ as 
\begin{equation}
    p(i|U, V, X) = P(\overrightarrow{e_u}, \overrightarrow{e_v}, \overrightarrow{e_x})
\label{eq:pFuncDemo}
\end{equation}
where N is the predefined number of items to be retrieved. ${\overrightarrow{e_v}}$ is the embedding of item v from set of items V.
Our framework outputs the probabilities for all the items, which represent how likely the specific user will engage with the items, and retrieves top-N candidate items.

The objective function for training our model is to maximize the following log-likelihood:
\begin{equation}
    l = \sum\limits_{(i, U, V, X)\in S} \log p(i|U, V, X)
\label{eq:pFuncDemo}
\end{equation}

We use the Adam optimizer to train our method.

\begin{figure*}
    \centering
    \includegraphics[width=\textwidth]{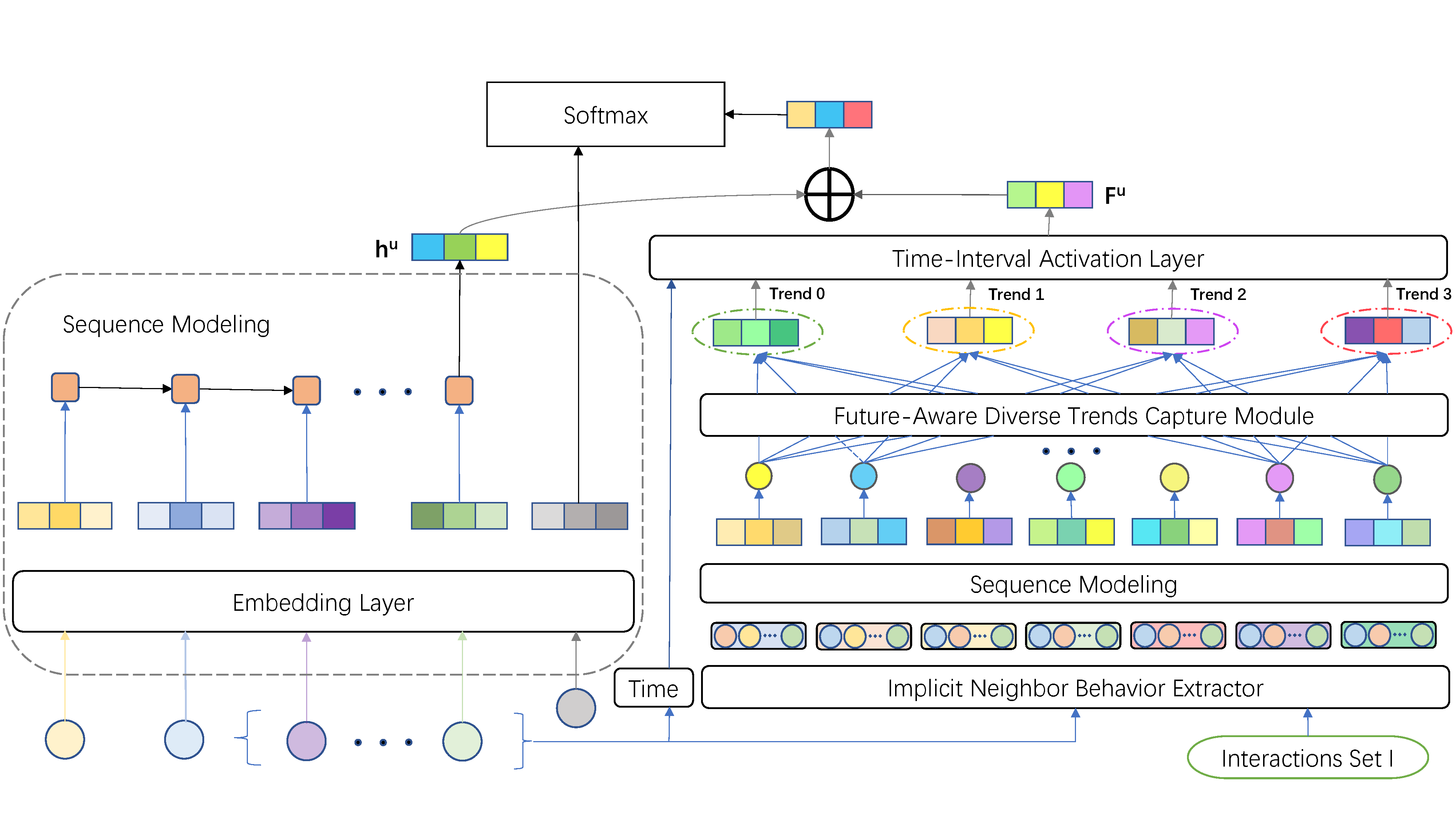}
    \caption{Network Architecture of FAT. The left part of FAT illustrates the network of our base model. The Base model takes user historical behaviors as inputs, and outputs user preference representation vectors ${h^u}$ for prediction decoder. The right part of FAT consists of an Implicit Neighbor Behavior Extractor, a sequence modeling module same as the left part which outputs relative future preference ${F^u}$, a diverse trends capture layer and a time-aware attention layer.}
    \label{fig:framework}
\end{figure*}

\subsection{Framework}
The overall structure of our proposed framework FAT is illustrated in Figure~\ref{fig:framework}, which is composed of a sequence modeling layer, an implicit neighbor behavior extractor, a diverse trends capture module, a time-aware attention layer and a final prediction layer. As the relative future sequence for current user is actually the history sequence for the neighbors, the future and history sequences can be modeled using shared parameters. Thus, we applied the sequential model for neighbors and users in the same manner. The framework takes the user historical interactions set ${X}$ and item features set ${F}$ as input. As for the extremely high dimension of item ID features, we adopt the widely-used embedding technique to embed these ID features into low-dimensional dense vectors. Here, we use ${s^{u}}$ and ${S^{u}}$ represent testing and training data of interactions sequence of user u respectively. For target item IDs from the set of ${S^{u}}$, embeddings are presented as ${\overrightarrow{e_{S^u}}}$.

The implicit neighbor behavior extractor constructs neighbors set for each user by filtering out the users that have interactions with the target items in the past, and then select the relative future sequences of each neighbor user. The target items can be selected from the user historical interaction ${X_u}$, for simplicity, we only choose the last one in the list, details are illustrated in Section \ref{sec:results}. The relative future behaviors are defined as the interacted items following the target item in the chronological order, aiming at representing dynamic preference for the user based on the intuition that the user tend to have similar preference trend as users with similar historical behaviors, and that the user can have diverse trends of preferences.

The diverse trends capture module is developed to obtain the neighbor centroids according to diverse motivation of specific interactions of the items. Then we learn high dimensional embeddings for the historical behaviors and future-aware diverse trends behaviors separately. Furthermore, the future sequence representation acquired by time-aware attention layer is concatenated with the historical behavior representation to generate the dynamic user preference representation vector. Finally we compute the user's preferences over different items from the pool by the prediction decoder. Each part will be elaborated in the following.

\subsection{Sequence Modeling}
\label{sec:seqMod}
We first computes user embeddings from user historical behaviors and then the diverse trends embeddings from the implicit neighbor relative future behaviors. To capture the dynamics of interaction sequences, we apply RNN to compute embeddings for users. The input of our sequence modeling module is the user historical behavior sequence or the relative future behavior sequences from the extracted neighbors, which contains a list of item IDs representing the user's interactions with items in time order. The number of item IDs is about billions, thus we adopt the widely-used embedding technique to embed these ID features into low-dimensional dense vectors, which significantly reduces the number of parameters and eases the learning process. The item IDs are fed into an embedding layer and transformed into item embeddings. To capture time-varying preferences of users, we then apply Recurrent Neural Network(RNN) to model the variable-length sequence data to compute the user embeddings. Particularly, we use Long Short-Term Memory cell as the basic RNN unit, which captures temporal dynamics. Each LSTM unit at time ${t}$ consists of a memory cell ${c_t}$, an input gate ${i_t}$, a forget gate ${f_t}$, and an output gate ${o_t}$. These gates are computed from previous hidden state ${h_{t-1}}$ and the current input ${x_t}$:
\begin{equation}
    [f_{t}, i_{t}, o_{t}] = sigmoid(W[h_{t-1}, x_t])
\label{eq:gates}
\end{equation}
The memory cell ${c_t}$ is updated by partially forgetting the existing memory and adding a new memory content ${I_t}$:
\begin{equation}
    I_{t} = tanh(V[h_{t-1}, x_t])
    c_{t} = f_t \odot c_{t-1} + i_t \cdot I_t
\label{eq:prediction}
\end{equation}
Once the memory content of the LSTM unit is updated, the hidden state at time step ${t}$ is given by:
\begin{equation}
    h_{t} = o_{t} \odot tanh(c_t)
\label{eq:prediction}
\end{equation}
At time step ${t}$, the new states of the user can be inferred as:
\begin{equation}
    h_{i,t}^{u} = LSTM(h_{i,t-1}^{u}, z_{i,t}^{u})
\label{eq:newState}
\end{equation}
where ${h_{i,t}^{u}\textbf{}}$ denote hidden states for the user. 
Note that LSTM can be replaced by other options, such as a Gated Recurrent Unit. In this paper, we select LSTM as it is a popular and general choice in previous works \cite{wu2017RRN,zhu2017timeLSTM,zhang2019gacoforrec,xu2019RCNN}.

Specifically, item embedding and user historical behavior embedding can be represented as ${E_V = \{\overrightarrow{\mathbf{e_v}}, v \in V\}}$ and ${E_X = \{\overrightarrow{\mathbf{h_x^u}}, x \in X_u\}}$, respectively. ${F_n}$ denotes the relatively future sequence of neighbors n.${E_N = \{\overrightarrow{\mathbf{f_x^n}}, n \in N_u, x \in F_n\}}$ represent the implicit similar users future behavior embedding, which is the input of the future-aware diverse trends capture module, output diverse motivations behind behaviors representing diverse trends of user dynamic preferences, and then through the attention layer, we obtain a single aggregated trend representation vector of a specific user.

Lastly, corresponding user historical behavior embedding and the trend behavior embedding are concatenated to form the user preference embedding ${E_u}$.

\subsection{Implicit Neighbor Behavior Extractor}
\label{sec:INBE}
Inspired by some works \cite{Guo2016socialInfluence,Felicio2016socialInfo}, which extract social relationships in absence of explicit social networks \cite{mukherjee2019ghostlink} , we compare the similarity among users via collaborative filtering to extract neighbor behaviors implicitly based on the historical interactions.

We adopt Pearson Correlation Coefficient \cite{breese2013empirical} as:
\begin{equation}
    s_{ij} = \frac{\sum\limits_{k\in I(i) \cap I(j)}(r_{ik} - \overline r_i) \cdot (r_{jk} - \overline r_j)}{\sqrt{\sum\limits_{k\in I(i) \cap I(j)}(r_{ik} - \overline r_i)^2} \cdot \sqrt{\sum\limits_{k\in I(i) \cap I(j)}(r_{jk} - \overline r_j)^2}}
\label{eq:pearsoncc}
\end{equation}
where ${I(i)}$ is a set of items that interacted/rated by user ${i}$, ${r_{ik}}$ and ${\overline r_i}$ represents the rate of user ${i}$ over item ${k}$ and the average rate of user ${i}$. The user similarity $s_{i}$ is ranging from ${[-1, 1]}$, and the similarity between users ${i}$ and ${j}$ is proportional to the value according to this definition. Following \cite{Hao2013implicitSocialRec}, we employ a mapping function ${f(x) = (x+1) / 2}$ to bound the range of PCC similarities into ${[0, 1]}$.

In the case of users with only one common item in history, PCC similarity gets $1$ when the users’ preferences over the common item are similar and $-1$ when not, which encourages diversity of neighbors while damaging the fairness of similarity calculation. To tackle this issue, we only kept less than ${20\%}$ of such neighbors to seek the balance. 

In addition to the PCC method, we also design a filter with simple schema to extract similar users. For each user, if the historical interactions ${I_u}$ is split into two pieces, ${\{S^{u}_{1:t}(t<|I_u|)\}}$ for training data, and ${\{s^{u}_{t+1:|I^u|}\}}$ for testing data, the item ${s^{u}_{K}}$ is defined as the last K target items, K could be any value less than or equal to ${|S^u|}$, while in practice ${K=1}$ can achieve good enough performance with simplicity, details would be illustrated in the Section 4.4 where we do an ablation of K. We extracted a list of users ${N=\{n_1, n_2, ..., n_{|N|}\}}$ from the item-user map using the target item as key, which stands for all the users who have interacted with the target item. Furthermore, we constructed the future sequence of each neighbor user ${u}$ relative to the target item ${s_{t'}}$ as:
\begin{equation}
    F_u = \{s_i, s_i \in I_n, R(s_i) \geq R(s_{t'})\}
\label{eq:neighborFutureSeq}
\end{equation}
where Timestamp is denoted as ${R}$ and ${s_{t'}}$ is the same item as ${s^{u}_{t} }$.

\subsection{Future-Aware Diverse Trends}
We argue that representing user neighbors by one representation vector can be a bottleneck for capturing diverse neighbors of users, because we have to compress all information related with diverse neighbors of users into one representation vector. Thus, all information about diverse neighbors of users is mixed together, causing inaccurate neighbor retrieval and then the inaccurate item retrieval for the matching stage. Instead, we adopt multiple representation vectors to express distinct neighbors of users separately. By this way, diverse neighbors of users are considered separately in the matching stage, enabling more accurate neighbor retrieval as well as the item retrieval for every aspect of reasons.

We utilize clustering process to group neighbors(represented by historical behaviors of user's diverse) extracted via previous multi-hop filter into several clusters. Neighbors from one cluster are expected to be closely related and collectively represent one particular aspect of user behaviors. Here, we design the multi-neighbor extractor layer for clustering historical behaviors and inferring representation vectors for resulted clusters.

Since the design of multi-neighbor extractor layer is inspired by MIND\cite{cen2020cmind}, which has already revisited essential basics of dynamic routing for representation learning in capsule network, we'll explain how our designed multi-neighbor extractor layer work based on it.

The objective of the multi-neighbor extractor layer is to learn representations for expressing properties of user behaviors as well as whether corresponding behaviors exist. The semantic connection between capsules and neighbor representations motivates us to regard the neighbor representations as neighbor capsules and employ dynamic routing to learn interest capsules from neighbor capsules. nevertheless, the original routing algorithm proposed for image data is not directly applicable for processing user neighbor data. So, we propose Neighbor-to-interest dynamic routing for adaptively aggregating user's neighbors into interest representation vectors, and it differs from original routing algorithm in three aspects.

Let ${e_{i}}$ be the capsule ${i}$ of the primary layer. We then give the computation of the capsule ${j}$ of the next layer based on primary capsules. We first compute the prediction vector as:
\begin{equation}
    \hat{e}_{j|i} = W_{ij}e_{i}
\label{eq:prediction}
\end{equation}
where ${W_{ij}}$ is a transformation matrix. Then the total input to the capsule ${j}$ is the weighted sum over all prediction vectors ${\hat{e}_{j|i}}$ as:

\begin{equation}
    s_{j} = \sum\limits_{i}c_{ij}\hat{e}_{j|i}
\label{eq:capsuleinput}
\end{equation}
where ${c_{ij}}$ are the coupling coefficients that are determined by the iterative dynamic routing process.

We use "routing softmax" to calculate the coupling coefficients using initial logits ${b_{ij}}$ as:
\begin{equation}
    c_{ij} = \frac{exp(b_{ij})}{\sum_{k}exp(b_{ik})}
\label{eq:couplingCoef}
\end{equation}
where ${b_{ij}}$ represents the log prior probability that capsule ${i}$ should be coupled to capsule ${j}$. To ensure short vectors and long vectors to get shrunk to almost zero length and a length slightly below 1. Then the vector of capsule ${j}$ is computed by:
\begin{equation}
    v_{j} = squash(s_j) = \frac{ {\left \| s_{j} \right \| }^{2}}{1 + {\left \| s_{j} \right \| }^{2}} \frac{s_j}{\left \| s_{j} \right \| }
\label{eq:capsuleVector}
\end{equation}
where ${s_j}$ is the total input of capsule ${j}$.

The output trend capsules of the user ${u}$ are then formed as a matrix ${V_u = [v_{1}, ..., v_{K}] \in \mathbb{R}^{d \times K}}$  for downstream task.

\subsection{Time-Aware Attention Layer}
For history sequence representation, we simply use the output of the sequence modeling layer given the input of user's history interactions list, which contains ${K}$ future potential sequence representations. Then we utilize the time-aware attention to activate the weight of diverse trends to capture the timeliness of each trend. Specifically, the attention function takes the interaction time of item ${i}$, the interaction time of trends and trend embeddings as the query, key and value respectively. We compute the final future sequence representation of user ${u}$ as:
\begin{equation}
    HF_u = Attention(\overrightarrow{T_{i}}, \overrightarrow{T_{tr}}, \overrightarrow{t_{u}}) = \overrightarrow{t_{u}} softmax(pow(\overrightarrow{T_{i}}, \overrightarrow{T_{tr}}))
\label{eq:timeInterval}
\end{equation}
where Attention denotes the attention function, ${T_{i}}$ represents the interaction time of item ${i}$, $[T_{tr}]$ represents the average interaction time of items related to the trend, ${\overrightarrow{t_{u}}}$ represents the embedding of the trend.

\subsection{Prediction}
After computing the trend embeddings from activated trends through time-aware attention layer, we concatenate it with the user historical behavior embedding to form a user preference embedding. Given a training sample ${u, i}$ with the user preference embedding and item embedding, we can predict the possibility of the user interacting with the item as

\begin{equation}
    p(i|U, V, X) = \frac{exp(\overrightarrow{e_u}^T \overrightarrow{e_i})}{\sum_{v \in V} exp(\overrightarrow{e_u}^T \overrightarrow{e_v})} 
\label{eq:predictFunc}
\end{equation}

\section{Experiments}
\label{sec:EXP}
In this section, we first cover the dataset and experimental settings. And then we conduct extensive experiments and in-depth analysis to verify the performance of FAT for recommendation.

\subsection{Dataset}
We used three large benchmark datasets. The statistics of the three datasets are shown in \ref{tab:staData}.
\begin{itemize}
\item {\verb|Amazon Books|}: This dataset contains product reviews and metadata from Amazon, including 142.8 million reviews spanning May 1996 - July 2014. It includes reviews (ratings, text, helpfulness votes), product metadata (descriptions, category information, price, brand, and image features), and links (also viewed/also bought graphs).
\item {\verb|Steam|}: This dataset contains more than 40k games from steam shop with detailed data including reviews and information about which games were bundled together.
\item {\verb|Movielens-1M|\cite{Harper2015TheMD}}: One of the currently released MovieLens datasets, which contains 1,000,209 movie ratings from 6,040 users across 3,900 movies. 

\begin{table}[h]
\caption{Statistics of the Datasets.}
\centering
\resizebox{\columnwidth}{!}{%
\begin{tabular}{l cc cc c}
\toprule
Dataset & users & items & interactions  \\
    \midrule
    Amazon Books   &459,133 & 313,966 & 8,898,041 \\ 
    Steam          & 2,567,538 &15,474 & 7,793,069 \\ 
    MovieLens-1M   & 6,040 & 3,416 & 999,611  \\
    \bottomrule
\end{tabular}%
}
    \label{tab:staData}
\end{table}

\end{itemize}

In each dataset, we partition user's interactions into training, validation and test set by the proportion of ${8:1:1}$. To avoid data sparsity, we filter out the users and items with only few interactions in our experiment. In the Movielens dataset, we keep users and items with at least 10 and 3 records respectively. In the Amazon Books dataset, we select users and items with at least 10 records each. 
In detail, we adopt a common setting of training sequential recommendation models. Let the behavior sequence of user ${u}$ be ${X_u = \{ s_1^u, s_2^u, ..., s_{|X_u|}^i \}}$. Each training sample uses the first ${k}$ behaviors of ${u}$ to predict the ${(k + 1)-th}$ behavior, where ${k = 1, 2, ..., |X_u|}$.

To evaluate, we randomly select an interacted item by the user as target item for each user, while the items interacted before the target item are collected as the user behaviors.

\begin{table*}[h]
\centering
    \caption{Model Performance on public datasets: Amazon books, Steam and MovieLens. FAT is our model and Base model is FAT without diverse trends. Here, we set K = 1 and T = 6}
\begin{tabular}{l cc cc cc cc cc cc}
\toprule
&\multicolumn{4}{c}{Amazon Books } & \multicolumn{4}{c}{Steam } & \multicolumn{4}{c}{MovieLens } \\
\midrule
Model & \multicolumn{2}{c}{Metrics@20} & %
    \multicolumn{2}{c}{Metrics@50} & \multicolumn{2}{c}{Metrics@20} & %
    \multicolumn{2}{c}{Metrics@50} & \multicolumn{2}{c}{Metrics@20} & %
    \multicolumn{2}{c}{Metrics@50}\\
\cmidrule(lr){2-3}\cmidrule(lr){4-5}\cmidrule(lr){6-7}\cmidrule(lr){8-9}\cmidrule(lr){10-11}\cmidrule(lr){12-13}
    & Recall & NDCG & Recall & NDCG  & Recall & NDCG & Recall & NDCG & Recall & NDCG & Recall & NDCG \\
    \midrule
    GRU4Rec       & 3.670 & 5.575 & 5.328 & 7.075  & 2.771 & 3.601 & 3.415 & 4.224 & 23.028 & 23.875 & 33.233 & 33.636\\ 
    YouTube DNN   & 3.933 & 5.703 & 6.612 & 7.623  & 2.812 & 3.711 & 3.667 & 4.373 & 23.676& 24.102 & 33.592 & 33.847\\ 
    MIND          & 4.102 & 5.933 & 6.638 & 7.830 & 3.213 & 4.331 & 3.671 & 4.591 & 24.750& 25.853 & 33.685 &  34.783\\ 
    ComiRec       & 4.853 & 6.185 & 7.203 & 8.120 & 3.464 & 4.142 & 3.977 & 4.792 & 24.883 & 25.896 & 33.955 & 34.984\\ 
    \midrule
    Base Model    & 3.126 & 4.912 & 4.872 & 6.721 & 2.481 & 3.812 & 3.217 & 4.123 & 23.101 & 23.879& 33.315 & 33.132\\ 
    FAT           & ${\mathbf{4.923}}$ & ${\mathbf{6.612}}$ & ${\mathbf{7.882}}$ & ${\mathbf{8.882}}$ & ${\mathbf{3.428}}$ & ${\mathbf{4.543}}$ & ${\mathbf{4.017}}$ & ${\mathbf{4.954}}$ & ${\mathbf{25.016}}$ & ${\mathbf{26.502}}$ & ${\mathbf{34.288}}$ & ${\mathbf{35.166}}$ \\ 
    \bottomrule
\end{tabular}
    \label{tab:modPer}
\end{table*}

\subsection{Evaluation Metrics}
To compare the performance of different models,we use \textbf{Recall@N} and \textbf{NDCG@N}(Normalized Discounted Cumulative Gain), where N is set to 20, 50 respectively as metrics for evaluation. In all these three metrics, a larger value implies better performance. Besides, we adopt per-user average for each metric.
\begin{itemize}
\item {\verb|Recall|}: Number of corrected recommended items divided by the total number of all recommended items.
\begin{equation}
    Recall@N = \frac{1}{|U|} \sum\limits_{u \in U} \frac{|\hat{I}_{u,N} \cap I_{u}|}{|I_u|}
\end{equation}
where ${\hat{I}_{u,N}}$ denotes the set of top-N recommended items for user u and ${I_u}$ is the set of testing items for user u.

\item {\verb|Normalized Discounted Cumulative Gain|}(NDCG): NDCG not only measures the percentage of correct recommended items, but takes the positions of correct recommended items into consideration.
\begin{equation}
    DCG@N = \frac{1}{|U|} \sum\limits_{u\in U} \sum\limits_{r\in R} \frac{\delta_{N}(r)}{log_{2}(i_{r} + 1)},
\end{equation}
\begin{equation}
    NDCG@N = \frac{DCG@N}{IDCG@N}
\end{equation}
where G denotes the ground-truth list. ${i_{r}}$ is the index of r in R. ${\delta_{N}(\cdot)}$ is an indicator function which returns 1 if item r is in top-N recommendation, otherwise 0. IDCG is the DCG of ideal ground-truth list which refers to the descending ranking of ground-truth list in terms of predicted scores. 

\end{itemize}

\subsection{Competitors}
\begin{itemize}
\item {\verb|GRU4Rec|\cite{hidasi2015gru4rec}}: A typical sequential recommendation baseline being the first to propose the usage of recurrent neural networks in recommendation systems.
\item {\verb|YoutubeDNN|\cite{paul2016youtubednn}}: One of predominant deep learning models based on collaborative filtering systems incorporating with text and image information which have been successfully applied under industrial scenario.
\item {\verb|MIND|\cite{li2019mind}}: A novel industrial applicable recommendation model to capture users' multi-interest.
\item {\verb|ComiRec|\cite{cen2020cmind}}: A novel controllable multi-interest framework which can be used in sequential recommendation.
\item {\verb|Base Model|}: We construct a base model of FAT by ignoring the diverse trends capture module and simply modeling user preferences from historical behaviors.

\end{itemize}

\subsection{Results}
\label{sec:results}
We have trained our model with Adam utilizing the TensorFlow distributed machine learning system using 4 replicas on a Nvidia GPU. The model performance for the sequential recommendation is shown in Table \ref{tab:modPer}. We run experiments to dissect the effectiveness of our recommendation model. We compare the performance of FAT with a baseline model of FAT and four state-of-the-art models: GPU4Rec, YouTube DNN, MIND and ComiRec. All these models are running on the three datasets introduced above: Amazon Books, Steam and MovieLens. According to the results shown in Table \ref{tab:modPer}, our model FAT obtain better performance on all evaluation metrics of all the tasks than other models.

As shown in Table \ref{tab:modPer}, Recall and NDCG for the dataset MovieLens of all the models are higher than that for dataset Amazon Books and Steam. It's caused by the unbalanced size between the datasets that size of the MovieLens is much smaller than the other two datasets.

\begin{table}[h]
\centering
    \caption{Model Performance of parameter sensitivity. T denotes the number of trends. T = 2 means relative future sequence from neighbors are clustered to two trends}
\resizebox{\columnwidth}{!}{%
\begin{tabular}{l cc cc cc }
\toprule
&\multicolumn{2}{c}{Amazon Books } & \multicolumn{2}{c}{Steam } & \multicolumn{2}{c}{MovieLens } \\
\midrule
Metrics@50
    & Recall & NDCG & Recall & NDCG  & Recall & NDCG  \\
    \midrule
    FAT(T = 2)    & 4.042 & 4.428 & 2.232 & 2.623 & 20.329 & 22.981 \\ 
    FAT(T = 4)    & 5.993 & 6.728 & 3.017 & 3.693 & 27.981 & 28.989 \\
    FAT(T = 6)    & ${\mathbf{7.882}}$ & ${\mathbf{8.882}}$ & ${\mathbf{4.017}}$ & ${\mathbf{4.954}}$ & ${\mathbf{34.288}}$ & ${\mathbf{35.166}}$\\
    FAT(T = 8)    & 6.982 & 7.842 & 3.107 & 4.125 & 31.973 & 32.887\\ 
    \bottomrule
\end{tabular}%
}
    \label{tab:parameterTrend}
\end{table}

Table \ref{tab:parameterTrend} reports the performance of our model FAT in different parameter setting by changing number of trends T. We list the performance result of our model for the three datasets setting T to 2, 4, 6 and 8. Our model achieves improvements on T = 6 over T = 4, which may caused by insufficient trends for the dataset. However, it did not show much importance when we change T from 6 to 8 showing clustering sequences to 8 trends in these datasets is redundant and to 6 trends is just suitable.

\begin{table}[h]
    \caption{Model Performance of Implicit Neighbor Behavior Extractor with ablation of K for the target items(setting T = 6). K = 3 means the last item, the second to last time and the third to last time are taken into consideration}
\centering
\resizebox{\columnwidth}{!}{%
\begin{tabular}{l cc cc cc }
\toprule
&\multicolumn{2}{c}{Amazon Books } & \multicolumn{2}{c}{Steam } & \multicolumn{2}{c}{MovieLens } \\
\midrule
Metrics@50
    & Recall & NDCG & Recall & NDCG  & Recall & NDCG  \\
    \midrule
    FAT(K = 1)    & 7.882 & 8.882 & 4.017& 4.954& 34.288 &35.166 \\ 
    FAT(K = 3)    & ${\mathbf{8.512}}$ & ${\mathbf{9.343}}$ & ${\mathbf{5.431}}$ & ${\mathbf{5.735}}$ & ${\mathbf{35.890}}$ & ${\mathbf{36.192}}$ \\
    FAT(K = 5)    & 8.482 & 8.912 & 5.117 & 5.654 & 35.588 & 35.894\\ 
    \bottomrule
\end{tabular}%
}
    \label{tab:parameterLastK}
\end{table}

Table \ref{tab:parameterLastK} compares the result of setting target item from the first to last item, the third to last item and the fifth to last one. The largest improvements appear on increasing K from 1 to 3. This demonstrates that by adding the number of target items, our model can capture more trend information and be more powerful to predict future sequences. Increasing target item number from 3 to 5 does not gain much improvement. This implies our model is efficient to capture much trend information by few historical items of the user.

\begin{table}[h]
    \caption{Model Recommendation Diversity with ${K = 4}$}
\centering
\begin{tabular}{l c c c }
\toprule
& Amazon Books & Steam & MovieLens \\
\midrule
Metrics@50
    & Diversity & Diversity & Diversity  \\
    \midrule
    GRU4Rec       & 36.783 & 40.648 & 20.875  \\ 
    YouTube DNN   & 38.604 & 42.831 & 23.654 \\ 
    MIND          & 39.967 & 44.984 & 27.502 \\ 
    ComiRec       & 42.915 & 45.947 & 28.961 \\
    \midrule
    Base Model  & 33.946 & 35.484 & 15.751  \\ 
    FAT         & ${\mathbf{43.591}}$ & ${\mathbf{46.653}}$ & ${\mathbf{29.274}}$ \\ 
    \bottomrule
\end{tabular}
    \label{tab:diversity}
\end{table}

Table \ref{tab:diversity} summarizes the recommend diversity performance of baseline models and our models on the three datasets.

The computational complexity of sequence layer modeling user and neighbors is $O(knd^2)$, where $k$ denotes the number of extracted neighbors, $n$ denotes the average sequence length and $d$ denotes the dimension of item’s representation. Capsule layer's computational complexity depends on kernel size and number of trends. Average time complexity of capsule layer scales $O(nTr^2)$, where $r$ denotes kernel size of capsule layer and $T$ denotes the number of trends. 
For large-scale applications, our proposed model could reduce computational complexity by two measures: (1)encode neighbors with a momentum encoder\cite{kaiming2020momentum}.(2)adopt a light-weight Capsule network.

\begin{figure}[htb]
    \centering
    \includegraphics[width=\linewidth]{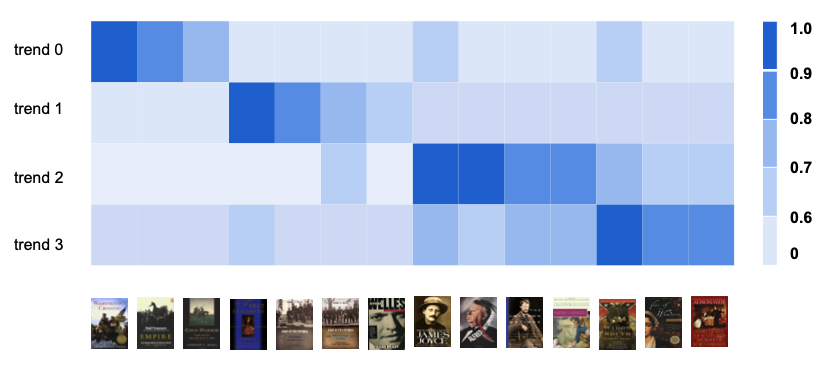}
    \caption{Heatmap of coupling factors for items recalled by each trend. Each item has the coupling factor on the corresponding trend. The color depth is proportional to the numerical value of the coupling factor.}
    \label{fig:heatmap}
\end{figure}

\begin{figure}[h]
    \centering
    \includegraphics[width=\linewidth]{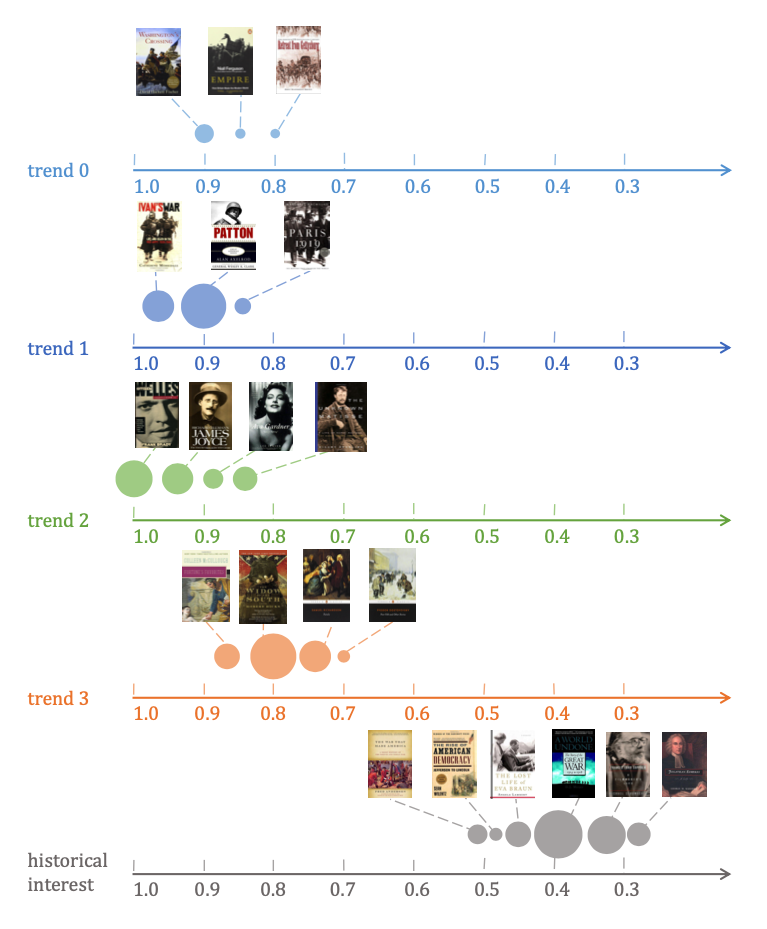}
    \caption{The distribution of items recalled by each trend (trend 0,1,2,3)and base model without trend(no trend). The coordinates indicate the recommendation level of items, where 1 signifies that the corresponding item is most recommended while 0 matches to the least recommended one. The radius of each circle is proportional to the number of items.}
    \label{fig:trendActivation}
\end{figure}

\subsection{Recommendation Diversity}
In addition to achieving high accuracy of recommendation, the diversity is also a critical part for user experience. Recommendation systems are trained to help users to select items which would be interesting to them without much historical interactions between the user and the items. Recommender systems tracks the interaction between the users and their selected items. This information is then processed to train the recommendation model which can not only recommend similar items but also recommend items of similar hidden connection.

Many authors have undertaken research developing new diversification algorithms \cite{Kwon2012ImprovingDiversity,boim2011diversification,noia2014propensity,prem2013EnhancingDiversity}. Our proposed module can learn the diverse trends of user preference and provide recommendation with diversity. Following \cite{cen2020cmind}, we use the following definition of individual diversity:

\begin{equation}
    Diversity@N = \frac{\sum_{j=1}^{N} \sum_{k=j+1}^{N} \delta(CATE(\hat{i}_{u,j}) \neq CATE(\hat{i}_{u,k}))}{N \times (N - 1) / 2}
\label{eq:diversityDef}
\end{equation}
where ${CATE}$ represents the category of the item. ${\hat{i}_{u}}$ denotes item recommended for user ${u}$, ${j}$ and ${k}$ represents the order of the recommended items. ${\delta(\cdot)}$ is an indicator function.

Table \ref{tab:diversity} shows the diversity of models on different datasets when we control the factor ${K=4}$. From the table, our module FAT achieve the optimum diversity indicating the recommendation it provide can effectively take neighbors' interests into account.

\subsection{Case Study}
\subsubsection{Coupling Factors}
The coupling factors between trends and items are proportional to the correspondence between them. In this section, we visualize these factors to show that the trend capture process is interpretable.

As shown in Figure \ref{fig:heatmap}, the coupling factors associated to the user randomly selected from Amazon Book dataset, where each row corresponds to one trend capsule and each column corresponds to one interaction after the selected target item. It shows that user X has interacted with 3 kinds of books (history, science, art) after interacting with the books of history category. Each of the future interactions has the max coupling factors on one trend capsule and forms the corresponding trend.

\subsubsection{Distribution}
We draw a trend distribution Figure \ref{fig:trendActivation} of recommended items recalled by each trend interest based on their similarity to the corresponding interest. Figure \ref{fig:trendActivation} shows the recommended item distribution for a user. X axis is the similarity and images are recommended item. The size of the circle demonstrate the recommended rate. As shown, the items recalled are correlated with trend interests.  

\section{Conclusions}
In this paper, we propose a novel Future-Aware Diverse Trend(FAT) framework to capture diverse trends of user preference dynamically. Our frame work leverages a neighbor behavior extractor to generate relative future interactions from similar users implicitly and utilizes diverse trends module to capture intrinsic varying dynamics of user preferences. To improve the expressive ability of trend representation, we utilize time-aware attention layer to make the duration between prediction time and target item interaction time choose which trend is more relative. Experimental results demonstrate that our models can achieve significant improvements over state-of-the-art models on three challenging datasets.
For the future, we plan to leverage multi-hop user-item graphs to address limited interaction issues and incorporate multi-behavior data into neighbors extraction to better model potential trends.

\bibliographystyle{ACM-Reference-Format}
\bibliography{main.bib}

\end{document}